\documentclass[conference]{IEEEtran}

\usepackage[letterpaper, left=0.62in, right=0.62in, bottom=1in, top=0.75in]{geometry}

\usepackage[table, dvipsnames]{xcolor}
\usepackage{blindtext}
\usepackage[T1]{fontenc}
\usepackage[utf8]{inputenc}
\usepackage{graphicx} 
\usepackage{amssymb} 
\usepackage{amsmath} 
\usepackage{makecell} 
\usepackage{dsfont} 
\usepackage{tikz}

\usepackage{pdfpages} 
\usepackage{rotating}
\usepackage{multirow}
\usepackage{cite}
\usepackage{balance}
\usepackage{lipsum}
\usepackage{hhline}
\usepackage{arydshln}

 \IEEEoverridecommandlockouts 

\usepackage{fancyhdr}
\usepackage[acronym,style=alttree,nomain,shortcuts,nonumberlist,nogroupskip]{glossaries}
\newacronym{aoi}{AoI}{age of information}
\newacronym{fcfs}{FCFS}{\textit{first-come-first-served}}
\newacronym{lcfs}{LCFS}{\textit{last-come-first-served}}
\newacronym{iid}{i.i.d.}{independent and identically distributed}

\newcommand{\figwa}{\columnwidth}

\DeclareMathOperator*{\argmaxA}{arg\,max}

\begin{document}

\title{It Is Rude to Ask a Sensor Its Age-of-Information: Status Updates Against an Eavesdropping Node}
\author{\IEEEauthorblockN{Laura Crosara, Nicola Laurenti, and Leonardo Badia}
\IEEEauthorblockA{Dept.\ of Information Engineering (DEI), University of Padova, Italy \\
email: \{laura.crosara.1@phd. , nicola.laurenti@ , leonardo.badia@ \}unipd.it}
}
\date{}
\maketitle
\thispagestyle{empty}
 \pagestyle{empty}

\begin{abstract}
We consider periodical status updates between a transmitter and a legitimate receiver, in the presence of an eavesdropper that is sometimes able to capture pieces of information.
We assume that, in the absence of such a threat, the connection between the transmitter and the receiver is controlled by the transmitter with the aim to minimize the age of information at the receiver's side.
However, if the presence of an eavesdropper is known, the transmitter may further tune the generation rate of status updates to trade off the age of information values acquired by the eavesdropper and the receiver, respectively.
To analyze this problem, we first propose a metric that combines both objectives according to a Bergson social welfare framework, and then we solve the problem of finding the optimal generation rate as a function of the probability of data capture by the eavesdropper.
This enables us to derive notable and sometimes counter-intuitive conclusions, and possibly establish an extension of the age of information framework to security aspects from a performance evaluation perspective.
\end{abstract}

\begin{IEEEkeywords}
Age of Information; Data acquisition; Modeling; Communication system security.
\end{IEEEkeywords}

\section{Introduction}
\label{introduction}

\Ac{aoi} has become a performance indicator adopted frequently to quantify the freshness of status updates from remote transmitters~\cite{yates2021age}. Many sensing applications require to track real-time content and, more than the average delay or the sheer throughput, their most important requirement is that the exchanged data be fresh.

Whenever a transmitter and receiver exchange status updates, the \ac{aoi} at the receiver is defined as~\cite{kaul2012real}
\begin{equation}
\delta(t) =  t - \sigma(t)
\label{eq:aoi}
\end{equation}
where $\sigma(t)$ is instant of generation of the last received update. As normally done in this kind of analysis \cite{kaul2011minimizing,wang2020preempt}, we consider zero propagation delay in the exchange, so time instants can be indifferently computed at the transmitter's or the receiver's side, and that whenever an update is generated at the transmitter's side, it always conveys fresh information \cite{munari2022role}. Resource limitations imply that updates can only be performed sporadically, obtaining a trend of $\Delta(t)$ that is linearly growing until an update is performed, which resets the \ac{aoi} to $0$.

Queueing systems are among the first models investigated under this lens, already in some seminal papers on the topic \cite{kaul2012real}. Even the study of a simple M/M/1 queue highlights the following beautiful conclusion. If we assume that the transmitter generates updates with exponentially \ac{iid} inter-generation times, with tunable rate $\lambda$, and the service of the queue, also a memory-less process, has rate $\mu$, so that the offered load is $\rho=\lambda/\mu$, the lowest \ac{aoi} is achieved at a certain intermediate value,
which is less straightforward than the delay- or throughput-optimizing conditions that are $\rho \to 0^{+}$ and $\rho \to 1^{-}$, respectively.
This reasoning can be extended to more complex systems by changing the queue policy \cite{crosara2022cost,champati2021minimum} or explicitly including other aspects such as medium access control
\cite{munari,badia2022game,yavascan2021analysis}. 

In the present paper, we want to add a new twist, by including a confidentiality objective 
 related to the adversarial presence of an eavesdropper.
To frame the problem in a classic setup, we consider a transmitter owned by Alice sending status updates to Bob, who plays the role of a legitimate receiver. Alice can tune the generation rate of update packets and the service procedure is according to a standard M/M/1 queue with \ac{fcfs} policy \cite{yates2012real}.
However, in addition to the aforementioned actors, an eavesdropper is present, aptly named Eve, who has the ability to  capture information sent by Alice to Bob. We assume that all updates from Alice are received by Bob, but each of them has probability $\beta \in [0,1]$ of being eavesdropped by Eve.

We further assume that Alice is aware of Eve's presence and knows the value of $\beta$. This changes the objective of the exchange from just sending fresh updates to Bob, to also including a \emph{further} goal of leaving only stale information to Eve.
Thus, the main contribution of this paper is a reformulation of the problem with a new objective function that chooses a point over the Pareto frontier of these two contrasting objectives according to Bergson's theory of social welfare \cite{bergson}.
This allows for an extension of the analytical framework to determine how 
the optimal load factor is influenced by Eve's probability of data capture. 

We discuss quantitative results and highlight important conclusions, such as the optimal generation rate being, under proper conditions, a decreasing function of the probability of data capture.
More in general, our investigation may set the basis for the extension of the age of information framework to security issues with analytical instruments.

The rest of this paper is organized as follows. In Section \ref{sec:rel}, we discuss models from the literature for \ac{aoi} of queuing systems, since our analysis piggybacks on them, and we also review the (actually few) efforts made to conjugate \ac{aoi} and security aspects. Section \ref{sec:mod} describes our proposed extension, from two different standpoints; first, we identify a trade-off between minimizing the \ac{aoi} of the legitimate receiver and maximizing that of the eavesdropper, and then we solve it through an entirely analytical framework. Section \ref{sec:res} presents numerical results. Finally, we conclude in Section \ref{sec:concs}.

\section{Related Work}
\label{sec:rel}
Many studies evaluate the \ac{aoi} in queuing systems, for various settings but especially based on classic memory-less systems with different disciplines \cite{moltafet2020average,costa2016age,crosara2021stochastic}.

The \ac{fcfs} M/M/1 queue presents a compelling behavior for what concerns its \ac{aoi}. On one hand, it is well known that its throughput is related to its stability, i.e., the arrival rate $\lambda$ and the service rate $\mu$ must satisfy $\rho =\lambda/\mu<1$, and a high throughput is achieved whenever $\rho$ approaches $1$. On the other hand, the delay is minimized when $\rho$ is close to $0$. 
The \ac{aoi} can be optimized by offering a traffic in an intermediate condition, even though the server is slightly biased towards being busy over being idle and so the optimal load factor $\rho$ is actually $\rho^\star \approx 0.531$ \cite{kaul2012real}. In other words, optimizing the \ac{aoi} in an M/M/1 queue implies seeking for non-aggressive management, where $\lambda$ is significantly lower than $\mu$, so there is already a self-limitation imposed to the data generation.

The quite elegant analytical results presented by Kaul and Yates in \cite{kaul2012real}, and subsequent contributions \cite{yates2018age}, are important sources of inspiration for the present work. In particular, the full expression of the average \ac{aoi} $\Delta= \mathbb{E}[\delta(t)]$ for an M/M/1 queue with \ac{fcfs} policy is \cite{kaul2012real}
\begin{equation}
    \Delta  = \lambda\big(\mathbb{E}[XT]+\mathbb{E}[X^2]/2\big)= \frac{1}{\mu}\left(1+\frac{1}{\rho} + \frac{\rho^2}{1-\rho} \right),
    \label{eqn:deltaonesource}
\end{equation}
where $X$ and $T$ are random variables equal to the interarrival time and system time of an update packet, respectively.

Some side remarks involve that there are substantially equivalent expressions, at least for what concerns the extensions meant in the present paper, to the cases of M/D/1, D/M/1, G/M/1, and so on, as well as with switching the discipline of the queue to \ac{lcfs}, adding preemption, and more \cite{champati2021minimum,moltafet2020average,costa2016age,talak2020age}.
For the purposes of our study, all of these evaluations can be considered equivalent, so we will just deal with the simpler M/M/1 queue.


Very few studies in the literature combine security and/or game theory with information freshness, and most of them just focus on mutual interference \cite{nguyen2018information} or intentional jamming \cite{vadori2015jamming,banerjee2022age}. The subject of confidentiality is rarely explored together with \ac{aoi}, which is surprising since many mission critical applications rely on timely exchanges, which an attacker may want to intercept, forge, or modify.
Paper \cite{wang2019joint} proposes to use \ac{aoi} as an integrated quality of service and security indicator to discriminate the validity of a hash key in a urban rail communication-based train control data communication systems. However, the \ac{aoi} is not used as a performance metric, but rather as a tool to improve secrecy. 
Similarly, \cite{jing2022joint} analyzes a generic Internet of Vehicles (IoV) network and designs a vehicle-assisted batch verification system.
Differently from \cite{wang2019joint}, they present a performance evaluation of \ac{aoi} as a quantitative indicator of security. 

In \cite{asheralieva2022optimizing}, the transmission system considers various scattered packets with some network coding connecting them, so that the receiver can decode the message after receiving $k$ packets out of $n$, but with the additional objective of preventing an eavesdropper from decoding that number of packets first. 

The closest contribution we can find to our proposed approach is \cite{chen2020secure}, where authors study the problem of maintaining information freshness under passive eavesdropping attacks. They consider a similar scenario, where a source sends its latest status to an intended receiver, while protecting the message from being overheard by an eavesdropper. Two \ac{aoi}-based metrics are defined to characterize the secrecy performance of the considered system. Also akin to our analysis, they obtain similar performance curves, on which they find the optimal data injection rate.
However, there are some notable differences with the present paper, which make our analysis simpler and more general. First of all, they consider a discrete time axis with stateful information, which allows for an optimization of the transmission schedule \cite{munari2022role,moltafet2020average}. We take a more basic approach where we  tune the arrival rate $\lambda$ of the queue. Since $\lambda$ is a continuous variable, our linear optimization is without any discretization effect.
Moreover, they consider a tradeoff between the \ac{aoi} performance at the intended receiver and at the eavesdropper, based on their difference. Instead, we investigate this from a wider perspective based on Bergson's theory of social welfare \cite{bergson} that allows to weigh the importance of contrasting the eavesdropper versus obtaining fresh information at the receiver.

Finally, combining conflicting objectives into a social welfare function according to Bergson's approach predates but is actually similar to the more well known contribution of Nash barganing \cite{nash1950bargaining}. Our specific choice corresponds to a product (that can be changed into a linear combination through logarithmic transformations) where exponential coefficients are tunable. The underlying point is that neither of the objectives can dominate over the other in a Pareto sense, but focusing on their product allows to identify a specific point on the Pareto frontier.

\section{Problem Definition} \label{sec:prob}
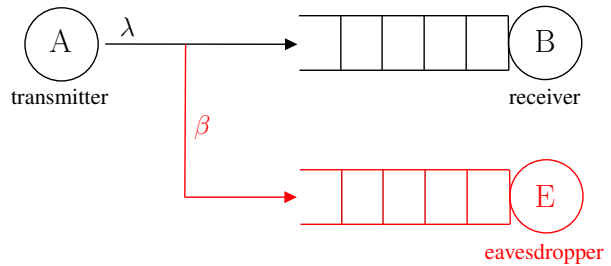
\begin{figure}
    \centering
     \resizebox{0.9\columnwidth}{!}{%
    \tikzset{every picture/.style={line width=0.75pt}} 

\begin{tikzpicture}[x=0.75pt,y=0.75pt,yscale=-1,xscale=1]

\draw    (220.83,16.5) -- (289.8,16.5) -- (370.19,16.5) ;
\draw    (221.26,55.45) -- (370.62,55.45) ;
\draw    (250.76,16.6) -- (250.76,56.02) ;
\draw    (280.19,16.02) -- (280.19,55.45) ;
\draw    (310.19,16.6) -- (310.19,55.17) ;
\draw    (340.19,16.31) -- (340.19,55.74) ;
\draw    (370.19,16.5) -- (370.62,55.45) ;
\draw   (370.4,35.98) .. controls (370.4,21.55) and (382.1,9.86) .. (396.52,9.86) .. controls (410.94,9.86) and (422.64,21.55) .. (422.64,35.98) .. controls (422.64,50.4) and (410.94,62.09) .. (396.52,62.09) .. controls (382.1,62.09) and (370.4,50.4) .. (370.4,35.98) -- cycle ;
\draw    (82.63,36.73) -- (216.83,36.91) ;
\draw [shift={(219.83,36.92)}, rotate = 180.08] [fill={rgb, 255:red, 0; green, 0; blue, 0 }  ][line width=0.08]  [draw opacity=0] (8.93,-4.29) -- (0,0) -- (8.93,4.29) -- cycle    ;
\draw   (25.4,36.98) .. controls (25.4,22.55) and (37.1,10.86) .. (51.52,10.86) .. controls (65.94,10.86) and (77.64,22.55) .. (77.64,36.98) .. controls (77.64,51.4) and (65.94,63.09) .. (51.52,63.09) .. controls (37.1,63.09) and (25.4,51.4) .. (25.4,36.98) -- cycle ;
\draw  [color={rgb, 255:red, 255; green, 0; blue, 4 }  ,draw opacity=1 ] (371.3,145.48) .. controls (371.3,131.05) and (383,119.36) .. (397.42,119.36) .. controls (411.84,119.36) and (423.54,131.05) .. (423.54,145.48) .. controls (423.54,159.9) and (411.84,171.59) .. (397.42,171.59) .. controls (383,171.59) and (371.3,159.9) .. (371.3,145.48) -- cycle ;
\draw [color={rgb, 255:red, 255; green, 0; blue, 4 }  ,draw opacity=1 ]   (140,36.71) -- (139.45,146.11) ;
\draw [color={rgb, 255:red, 255; green, 0; blue, 4 }  ,draw opacity=1 ]   (221.5,126.5) -- (290.47,126.5) -- (370.86,126.5) ;
\draw [color={rgb, 255:red, 255; green, 0; blue, 4 }  ,draw opacity=1 ]   (221.93,165.45) -- (371.29,165.45) ;
\draw [color={rgb, 255:red, 214; green, 0; blue, 4 }  ,draw opacity=1 ]   (251.43,126.6) -- (251.43,166.02) ;
\draw [color={rgb, 255:red, 214; green, 0; blue, 4 }  ,draw opacity=1 ]   (280.86,126.02) -- (280.86,165.45) ;
\draw [color={rgb, 255:red, 214; green, 0; blue, 4 }  ,draw opacity=1 ]   (310.86,126.6) -- (310.86,165.17) ;
\draw [color={rgb, 255:red, 255; green, 0; blue, 4 }  ,draw opacity=1 ]   (340.86,126.31) -- (340.86,165.74) ;
\draw [color={rgb, 255:red, 255; green, 0; blue, 4 }  ,draw opacity=1 ]   (371.19,126.48) -- (371.27,133.98) -- (371.62,165.43) ;
\draw [color={rgb, 255:red, 255; green, 0; blue, 4 }  ,draw opacity=1 ]   (216.4,146.08) -- (139.4,146.08) ;
\draw [shift={(219.4,146.08)}, rotate = 180] [fill={rgb, 255:red, 255; green, 0; blue, 4 }  ,fill opacity=1 ][line width=0.08]  [draw opacity=0] (8.93,-4.29) -- (0,0) -- (8.93,4.29) -- cycle    ;

\draw (387.5,25.5) node [anchor=north west][inner sep=0.75pt]  [font=\LARGE]  {$\mathrm{B}$};
\draw (41.5,25.5) node [anchor=north west][inner sep=0.75pt]  [font=\LARGE]  {$\mathrm{A}$};
\draw (387.7,135.5) node [anchor=north west][inner sep=0.75pt]  [font=\LARGE]  {$\mathrm{\textcolor[rgb]{1,0,0.02}{E}}$};
\draw (352.07,178) node [anchor=north west][inner sep=0.75pt]  [font=\normalsize] [align=left] {{\fontfamily{ptm}\selectfont {\large \textcolor[rgb]{1,0,0.01}{eavesdropper}}}};
\draw (14,67) node [anchor=north west][inner sep=0.75pt]  [font=\normalsize] [align=left] {{\fontfamily{ptm}\selectfont {\large transmitter}}};
\draw (369.33,67) node [anchor=north west][inner sep=0.75pt]  [font=\normalsize] [align=left] {{\fontfamily{ptm}\selectfont {\large receiver}}};
\draw (144.67,86.07) node [anchor=north west][inner sep=0.75pt]  [font=\Large]  {$\textcolor[rgb]{1,0,0.01}{\beta }$};
\draw (91.67,16) node [anchor=north west][inner sep=0.75pt]  [font=\Large]  {$\lambda $};

\end{tikzpicture}}
    \caption{Queuing system with a transmitter (A), a legitimate receiver (B), and an eavesdropper (E).}
    \label{fig:model}
\end{figure}
We consider a system as depicted in Fig. \ref{fig:model}, where a transmitter (Alice) sends status updates to a receiver (Bob). Alice can tune the generation rate of update packets and the service procedure is according to an \ac{fcfs} M/M/1 queue. We add a twist to this scenario adding an eavesdropper (Eve), that may capture data packets sent by Alice to Bob. 

In the absence of Eve, Alice objective would be to minimize the \ac{aoi} at Bob's receiver, to keep the information available to Bob as fresh as possible. However, if the presence of Eve is known, Alice may adjust the generation rate of status updates to increase the \ac{aoi} at Eve's receiver. Therefore, Alice seeks for a tradeoff between two objectives, i.e., minimizing the \ac{aoi} available to Bob and maximizing the \ac{aoi} at Eve's side.

A typical real-world scenario that could be cast into our system is represented, for instance, by a open communication environment, which makes wireless transmissions more vulnerable than wired communications to malicious attacks \cite{zou2016survey,zhu2017secure}. In particular, an eavesdropper can manage to intercept data whenever Alice and Bob cannot establish a secure communication channel.
Tactical networks \cite{thanh2019efficient} are also an important application for our analysis.

To sum up, we are going to address the following points.
\begin{enumerate}
    \item Define an appropriate confidentiality-aware objective function, which takes into account the two contrasting purposes of Alice, namely minimizing the average \ac{aoi} at Bob's receiver, while keeping the average \ac{aoi} at Eve's receiver as large as possible
    \item Find the optimal generation rate of update packet for Alice, according to the objective function above
    \item Show and discuss quantitative results, highlighting
    counter-intuitive conclusions, considering different scenarios and system parameters.
\end{enumerate}

\section{Analytical Model} 
\label{sec:mod}
We consider the system described in Section \ref{sec:prob}, where a transmitter (Alice) sends periodical update packets to a legitimate receiver (Bob) through a \ac{fcfs} M/M/1 packet queue. Each update transmitted by Alice carries new information that resets the \ac{aoi} at Bob's side.
Alice generates packets according to a Poisson process of rate $\lambda$ and service time of Bob's queue is exponentially distributed with rate  $\mu$, providing an offered load $\rho = {\lambda}/{\mu}$. 
It is not restrictive to normalize Bob's service capacity as $\mu = 1$, so that $\lambda=\rho$; otherwise, all the results can be rescaled by a factor $\mu$.
We assume that the channel between Alice and Bob is error-free, so that every update packet sent by Alice is correctly received by Bob, although in this basic framework it would be possible to account for erasures of status updates by simply modifying $\rho$ accordingly.

Moreover, we consider the presence of an eavesdropper, referred to as Eve (E), which attempts to capture the information exchanged between the transmitter and the receiver. We assume that each update packet transmitted by Alice is independently eavesdropped by Eve according to an \ac{iid} statistics, with eavesdropping probability $\beta\in[0,1]$. Consequently, we can consider that a fraction $\beta$ of the transmitted packets are received also by Eve. According to the thinning property \cite{principlesofcomm}, packets arrival at Eve's queue follow a Poisson process with rate $\beta\lambda$.
Akin to Bob, Eve enqueues her packets in a \ac{fcfs} M/M/1 queue with service rate $\mu$, equal to that of Bob. The load factor in the channel between Alice and Eve is $\rho_\mathrm{E} = {\beta\rho}$.

\subsection{Confidentiality Aware Objective Function}\label{sec:secmet}
In our scenario, Alice is assumed to be the only intelligent agent, since she can choose her transmission rate $\lambda$, while Eve and Bob are passive entities. We further assume that Alice is aware of Eve's presence and knows the value of $\beta$. 
In a scenario where no eavesdropper is present, the purpose of the transmitter will be to tune the value of $\rho$ to obtain an \ac{aoi} value at the legitimate receiver Bob that is as small as possible. However, the presence of an eavesdropper who captures a fraction of the transmitted packets implies that Alice wants the information available to Eve to be as old as possible, in addition to minimizing Bob's \ac{aoi}. Therefore, Alice has two competing objectives described by the utility functions
\begin{equation}
    u_1(\rho) = \frac{1}{\Delta_\mathrm{B}(\rho)}\,,\;\;\;\;\;\;\; u_2(\rho) =\Delta_\mathrm{E}(\rho),
    \label{eqn:utilities}
\end{equation}
where $\Delta_\mathrm{B}(\rho) = \mathbb{E}[\delta_\mathrm{B}(t)]$ and $\Delta_\mathrm{E}(\rho)= \mathbb{E}[\delta_\mathrm{E}(t)]$ represent the expected \ac{aoi} at Bob's and Eve's receivers, respectively.

From Alice's perspective, it is beneficial to increase either of these utilities, or both. However, they are contrasting objectives as is clear from the following reasoning. Indeed, the values of $\delta_\mathrm{B}(t)$ and $\delta_\mathrm{E}(t)$ increase until Alice generates a data packet. When a new data packet is transmitted, two situation can occur: (i) the packet is received by both Bob and Eve, this happens with probability $\beta$. In this case, at the current time instant $\delta_\mathrm{B}(t)$ and $\delta_\mathrm{E}(t)$ are reset to zero; (ii) the packet is received only by Bob, this happens with probability $1-\beta$. In this case, at the current time instant only $\delta_\mathrm{B}(t)$ is reset to zero while $\delta_\mathrm{E}(t)$ continues to increase. This means that, whenever $\delta_\mathrm{B}(t)$ is lowered, $\delta_\mathrm{E}(t)$ can decrease too, since Eve's capture of data cannot be controlled or forecast by Alice.

Thus, to maximize the two competing utilities of (\ref{eqn:utilities}), we reformulate the problem defining a new objective function that sets a precise value on Pareto frontier created by $u_1$ and $u_2$, i.e., the set of values for which $u_1$ cannot be increased without lowering $u_2$, or vice versa. This choice is made following Bergson's approach \cite{bergson}, where we set an ultimate objective function $f$ to be a weighted product between the two utilities $u_1$ and $u_2$, which is a modified Nash bargaining solution \cite{nash1950bargaining}
\begin{equation}
    f(\rho)=[u_1(\rho)]^{a+1} u_2(\rho)=\frac{\Delta_\mathrm{E}(\rho)}{[\Delta_\mathrm{B}(\rho)]^{a+1}},
\end{equation}
with $a\in(0,+\infty)$ being a parameter that controls the trade-off between $u_1$ and $u_2$. 
Note that in the choice of the exponent of $u_1$ we must assume that this objective cannot be eliminated; otherwise, we would reach a trivial allocation where Alice never updates. This would consistently obtain a very high $\Delta_\mathrm{E}(\rho)$ but would also have $\Delta_\mathrm{B}(\rho)$ to grow indefinitely, which goes against the motivation of the setup in the first place. Thus, the objective of delivering fresh data to Bob cannot be avoided and the exponent in the trade-off must be greater than or equal to $1$.
Hence, we write it as $a+1$, where the larger $a$, the more important $u_1$ versus $u_2$ in the trade-off. Moreover, $a \to + \infty$ corresponds to ignoring the presence of Eve, while $a \to 0^+$ means that the threat of the eavesdropping receives the highest importance, and Alice just wants to minimize the ratio $\Delta_\mathrm{B}(\rho)/\Delta_\mathrm{E}(\rho)$ instead of $\Delta_\mathrm{B}(\rho)$ itself. Therefore, the specific choice of $a$ governs the selection of the optimal point in the Pareto frontier.


\subsection{Optimal Offered Load}
The full expressions for $\Delta_\mathrm{B}(\rho)$ and $\Delta_\mathrm{E}(\rho)$ when $\mu=1$ can be computed from (\ref{eqn:deltaonesource}) 
as
\begin{equation}
    \Delta_\mathrm{B}(\rho) =
    1+\frac{1}{\rho} + \frac{\rho^2}{1-\rho}\,,
    \label{eqn:deltaB}
\end{equation}
for the legitimate channel between Alice and Bob, and
\begin{equation}
    \Delta_\mathrm{E}(\rho) = 
    1+\frac{1}{\beta\rho} + \frac{\beta^2\rho^2}{1-\beta\rho}\;.
    \label{eqn:deltaE}
 \end{equation}
for the eavesdropper channel between Alice and Eve.
The optimal offered load $\rho$ maximizing the objective $f(\rho)$ is
\begin{equation}
\begin{split}
    &\rho^\star = \argmaxA_\rho f(\rho)=  \argmaxA_\rho \frac{\Delta_\mathrm{E}(\rho)}{[\Delta_\mathrm{B}(\rho)]^{a+1}} \\
    &\,=\argmaxA_\rho \frac{1+\frac{1}{\beta\rho} + \frac{\beta^2\rho^2}{1-\beta\rho}}{\left(1+\frac{1}{\rho} + \frac{\rho^2}{1-\rho}\right)^{a+1}}\\
    &\,= \argmaxA_\rho \frac{(\beta^3\,\rho^3-\beta^2\,\rho^2+1)\rho^{a}\,(\rho-1)^{a+1}}{\beta\,\left(\beta\,\rho-1\right)\,{\left(\rho^3-\rho^2+1\right)}^{a+1}}
    \label{eqn:objfun}
\end{split}
\end{equation}
Equation (\ref{eqn:objfun}) can be solved by computing the derivative of $f(\rho)$. It is worth noting that, when $\beta\to 0^+$, the derivative $f'(\rho)$ approaches
\begin{equation}
    f'(\rho) \to \frac{\,g(\rho)(\rho\,-\rho^2)^a}{\rho\,\beta\,\left(\rho^3-\rho^2+1\right)^{a+2}}\,,
\end{equation}
where $g(\rho)$ is the 4-th degree polynomial
\begin{equation}
    g(\rho)=(a+2)(\rho^4-2\rho^3+\rho^2)-(2a+1)\rho+a.
    \label{eqn:g}
\end{equation}
Therefore, when $\beta\to 0^+$, the optimal load factor at the limit is obtained as the only real solution of $g(\rho)=0$ in the interval $(0,\mu)$.
The function $g(\rho)$ is continuous in the interval $(0,\mu)$ and it takes value of opposite sign at the boundaries
\begin{align}
    g(0) &= a\,>0 ,\\
    g(\mu) &=-(a+1) <0.
\end{align}
Therefore, according to the intermediate value theorem, a real value $\tilde{\rho}\in(0,1)$ such that $g(\tilde{\rho})=0$ must exist. Moreover, the first order derivative of $g(\rho)$ is
\begin{equation}
    g'(\rho)=2\rho(a+2)(\rho-1)(2\rho-1)-2a-1,
    \label{eq:asy}
\end{equation}
which is negative for every $\rho\in(0,1)$. Consequently, the solution $\tilde{\rho}$ is unique and can be found numerically. For example, in the case of $a=1$, we have
\begin{equation}
    3(\rho^2-2)(\rho^2+1)+1=0\,,
\end{equation}
and the solution is found at $\rho \approx 0.389$.

\begin{figure}
    \centering
    \includegraphics[width=1\figwa]{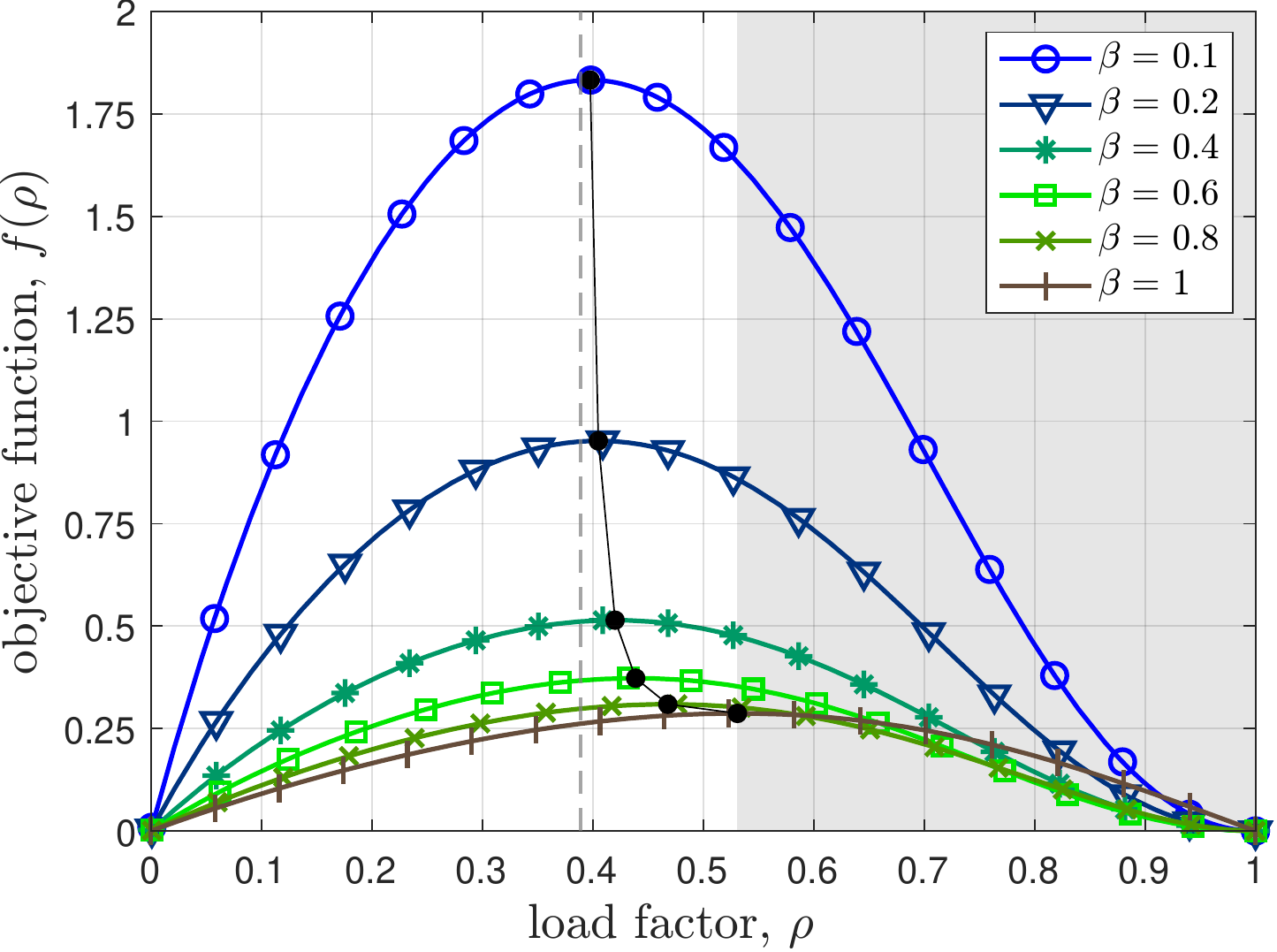}
    \caption{Objective $f(\rho)$, as a function of the load factor $\rho$, for different values of eavesdropping probability $\beta$, with weight $a=1$. The black line connects the maximizing points $\rho^\star$. The dashed black line reports $\rho= 0.389$.}
    \vspace{0.2cm}
    \label{fig:1}
\end{figure}
\section{Numerical Results} 
\label{sec:res}

We present quantitative evaluations to express the consequences of the derivations above.
The scenario considered includes a transmitter (Alice) and a receiver (Bob), whose communication is intercepted by an eavesdropper (Eve). Eve independently intercepts data packets with probability $\beta$. For sake of normalization, we consider both Bob's and Eve's service capacities to be $\mu=1$.
We will discuss how the optimal load factor $\rho^\star$, obtained maximizing the objective function $f(\rho)$ in (\ref{eqn:objfun}), is influenced by Eve's probability of data capture $\beta$ and the trade-off parameter $a$.

If Eve does not intercept any packet, i.e. $\beta=0$, we expect $\rho^\star = 0.531$, which is the \ac{aoi} minimizing value for the load factor with normalized service capacity \cite{kaul2012real}. When packets are eavesdropped with \ac{iid} probability $\beta>0$, we expect that the optimal load factor decreases, therefore $\rho^\star \leq 0.531$ for any value of $\beta$. For this reason, in all the results that follow, the areas corresponding to $\rho^\star > 0.531$ are shaded.

Fig. \ref{fig:1} shows the objective function $f(\rho)$, as a function of $\rho$ for different values of $\beta$ when $a=1$. The black line connects the maximum point of each curve, reached when $\rho = \rho^\star$, while the dashed black line reports the value $\tilde{\rho}= 0.389$. 
First of all, we note that the curves are bell-shaped with a very pronounced maximum when $\beta$ is small. When $\beta$ rises, the curves get flatter, this happens because, when $\beta$ tends to $1$, the two functions $\Delta_\mathrm{B}$ and $\Delta_\mathrm{E}$ get closer, and Alice has narrower margins to reach her objectives.
When $\rho=1$, all the curves go to zero. 
As the black line in Fig.\ \ref{fig:1} shows, the value of $\rho^\star$ tends to $0.531$ as $\beta$ increases, and decreases with $\beta$, tending towards a vertical asymptote at $\rho<0.531$, displayed as the black dashed line in the figure, whose numerical value is the solution of (\ref{eq:asy}). 
For the specific case of this figure where $a=1$, the asymptotic value shown by the vertical dashed line is $\tilde{\rho}= 0.389$.

\begin{figure}
    \centering
    \vspace{0.25cm}
    \hspace{-0.04cm}
    \includegraphics[width=0.98\figwa]{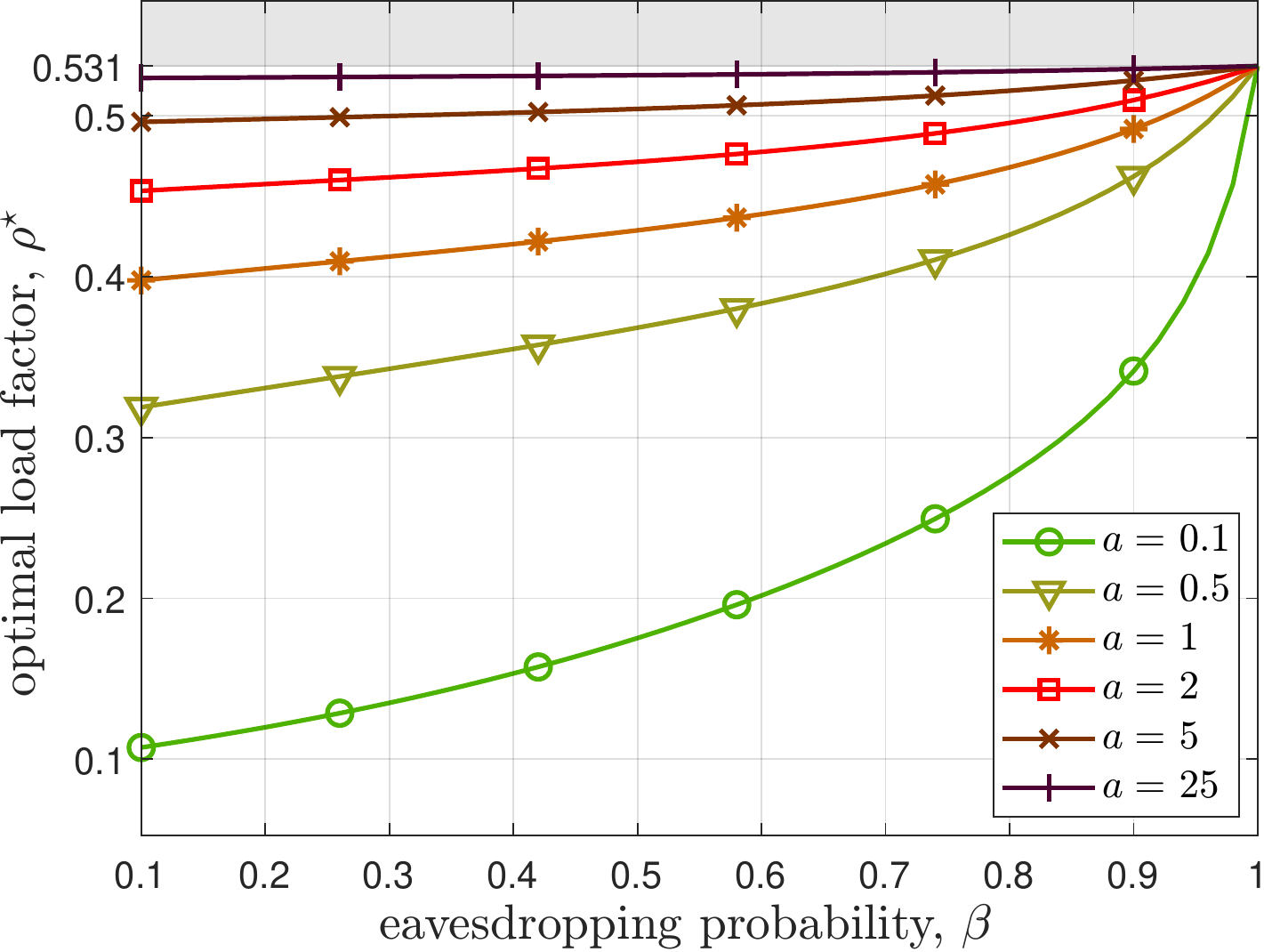}
    \caption{Optimal load factor $\rho^\star$, as a function of capture probability $\beta$, for different values of weight $a$.}
    \label{fig:2}
\end{figure}
Interestingly, the lower $\beta$, the lower $\rho^\star$, which, at first glance, may seem counter intuitive, yet this behavior has the following explanation.
If $\beta$ tends to $1$ the eavesdropper often intercepts the packets transmitted by Alice, so the only sensible objective for Alice is to keep $\Delta_\mathrm{B}$ low, which is achieved by choosing $\rho = 0.531$. 
If $\beta$ decreases, the second objective takes over, and Alice transmits less frequently, choosing $\rho<0.531$, to prevent Eve from intercepting. 
Above all, if $\Delta_\mathrm{B}$ is low and $\Delta_\mathrm{E}$ high, Alice should wait before transmitting a new packet because the effect can be to reset both $\Delta_\mathrm{B}$ and $\Delta_\mathrm{E}$. As a side note, in our analysis Alice only chooses the transmission rate $\lambda$, and she does not perform a real-time optimization based on the instantaneous values of the $\Delta_\mathrm{B}$ and $\Delta_\mathrm{E}$. Yet, it is expected that in a stateful optimization \cite{munari2022role,chen2020secure} (left for future research) this phenomenon will be seen with even more clarity.

Fig.\ \ref{fig:2} shows the optimal load factor $\rho^\star$ as a function of $\beta$, for different values of $a$.
One can see that the optimal value $\rho^\star$ approaches zero when the values of $\beta$ and $a$ are low. In other words, if the main objective for Alice is to have a large ratio of Eve's AoI versus Bob's, and Eve is rarely capable of eavesdropping data, the best strategy for Alice is also to update very rarely. This means that in Fig.\ \ref{fig:1} the dashed vertical grey line would move to the left as $a$ decreases.
Conversely, when the value of $a$ rises, $\rho^\star$ tends to $0.531$ for every $\beta$, thus
\begin{equation}
    \lim_{a\to+\infty} \rho^\star =  0.531\,,\;\;\forall \beta\in[0,1].
\end{equation}
Hence, the black dashed line in Fig. \ref{fig:1} would move to the left as $a$ increases.
When $\beta = 1$,  $\rho^\star=0.531$ for all $a> 0$.

\begin{figure}    
    \centering
    \vspace{0.2cm}
    \hspace{0.4cm}
    \includegraphics[width=0.98\figwa]{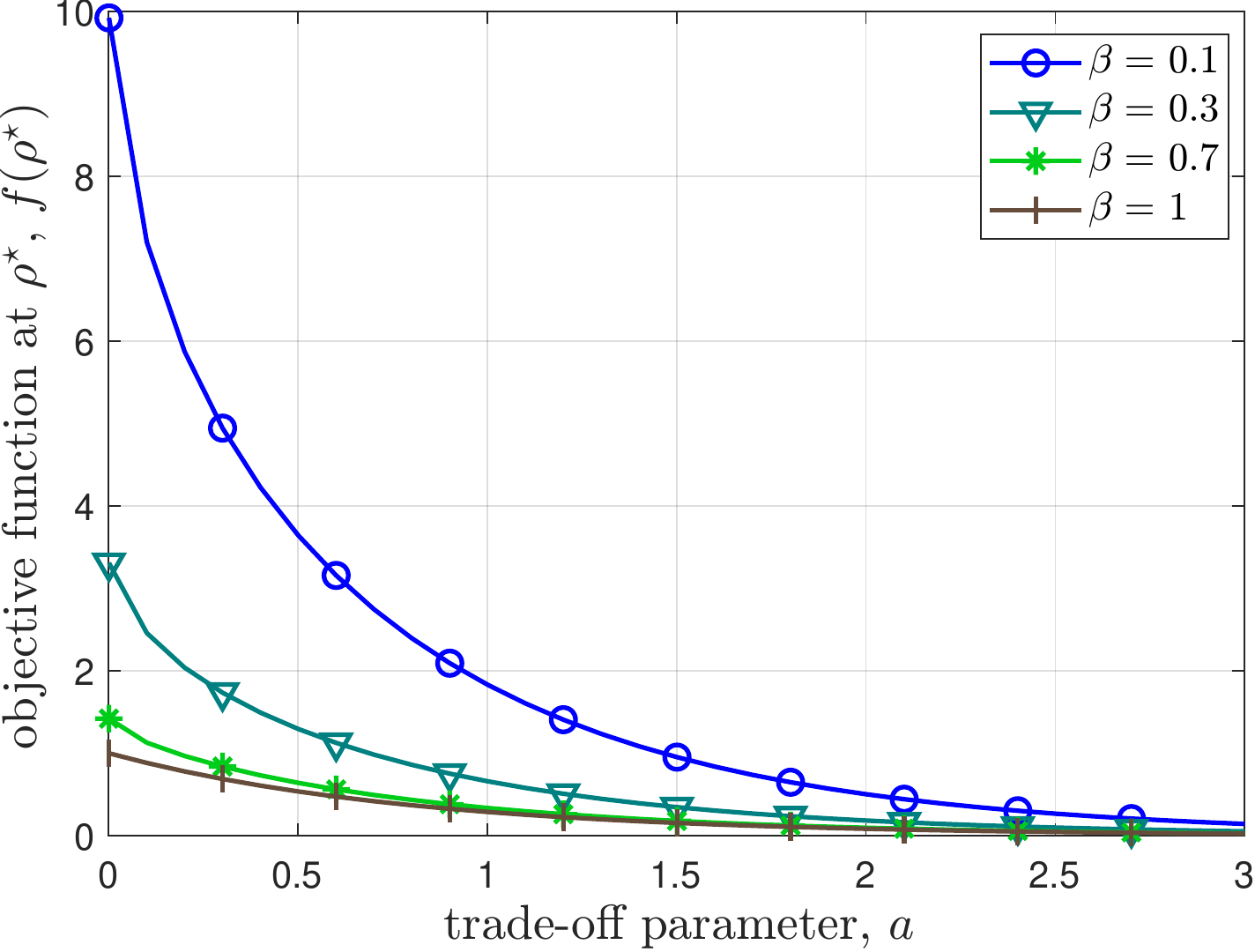}
    \caption{Objective function $f(\rho)$ evaluated at the optimal load factor $\rho^\star$, as a function of weight $a$, for different eavesdropping probabilities $\beta$.}
    \label{fig:3}
\end{figure}
Fig.\ \ref{fig:3} shows the objective function $f(\rho)$ evaluated at the optimal load factor $\rho^\star$, as a function of $a$, for different values of $\beta$. 
We note that when $a\to 0^+$, the value of the objective function at the optimal point $f(\rho^\star)$ tends to $1/\beta$ for every value of $\beta$, i.e.,
\begin{equation}
    \lim_{a\to0^+} f(\rho^\star) = \lim_{a\to0^+} \frac{\Delta_\mathrm{E}(\rho^\star)}{[\Delta_\mathrm{B}(\rho^\star)]^{(a+1)}} = 1/\beta,
\end{equation}
whereas, when $a\to + \infty$, $f(\rho^\star)\to 0^+$ for all values of $\beta$. 

\begin{figure}
    \centering
    \vspace{0.8cm}
    \hspace{-0.06cm}
    \includegraphics[width=\figwa]{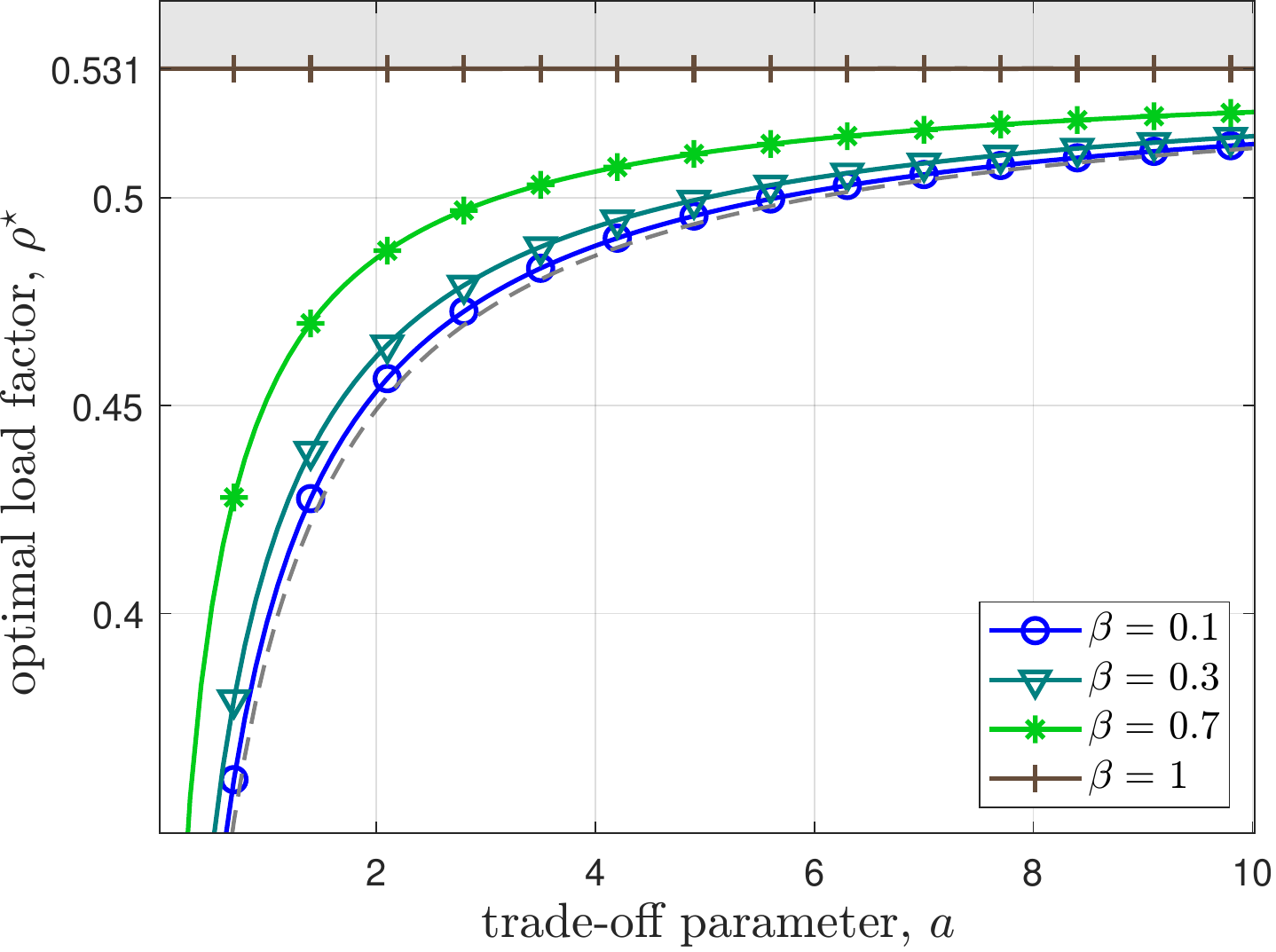}
    \caption{Optimal load factor $\rho^\star$, as a function of weight $a$, for different eavesdropping probabilities $\beta$. The black dashed line is the limit for $\beta \to 0^{+}$.}
    \label{fig:4}
\end{figure}
Fig.\ \ref{fig:4} shows the optimal load factor $\rho^\star$, as a function of $a$, for different values of $\beta$. 
For all the curves, the value of $\rho^\star$ moves toward $0$ when $a$ tends to zero, i.e.,
\begin{equation}
    \lim_{a\to0^+} \rho^\star = 0\,,\;\;\forall \beta\in[0,1),
\end{equation}
for every value of $\beta$, provided it is less than $1$ (whereas for $\beta=1$ the curve degenerates in a constant optimal choice of $\rho^\star=0.531$). 
Moreover, we also plot a black dashed line to represents the limit for $\beta \to 0^{+}$.
Notably, all curves with a relatively small (but not necessarily infinitesimal) values of $\beta$, such as $\beta=0.3$ in the figure, approach this asymptotic trend quite closely, thereby implying that for a low eavesdropping probability, the optimal behavior of the transmitter is always the same, and ultimately determined by the sole value of $a$, i.e., the level of importance attributed to one objective versus the other. We believe that this may lead to interesting conclusions about the optimal transmission policy for fresh status updates in the presence of an eavesdropper whenever the success rate of data capturing is relatively low, even in the case it is not accurately known.

\section{Conclusions} 
\label{sec:concs}
We analyzed a scenario of status updates between a transmitter and a legitimate receiver, considering also the presence of an eavesdropper that is sometimes able to intercepts data packets. 
For this purpose, we leveraged existing analytical results for queuing systems, where the \ac{aoi} is computed as a function of the load factor. 

We assume that the transmitter is aware of the eavesdropper and wants to set an efficient data injection rate
that simultaneously achieves low AoI at the intended receiver but keeps the information of the eavesdropper stale.
To analyze this problem, we proposed to combine both objectives according to a Bergson social welfare framework, then we solved the problem of finding the optimal load factor as a function of the probability of data capture by the eavesdropper.

The main conclusion is that, in order to account for this additional objective of leaking only stale information to the eavesdropper, the transmitter has to decrease its data generation rate, lowering the load factor.
Especially, if the predominant goal of the transmitter is to keep the eavesdropper at bay, the load factor tends to zero even for small values of the probability of data capture by the eavesdropper.
More in general, the present framework can be used as an adjustable approach for different cases of interests in practical contexts.

Envisioned extensions of the present paper include the analysis of an optimized schedule with stateful information \cite{munari2022role}, also investigating the costs for tracking the eavesdropper and detecting whether data was actually captured.
Moreover, a natural follow-up would be to consider the analysis of this adversarial setup from a game theoretic standpoint \cite{perin2021adversarial}, with an eavesdropper that is able to strategically regulate the data capture probability.
Finally, an extension to pervasive data networks, especially for what concerns the scalability of the analysis, is key to bring the present investigation in the context of future generation communication systems \cite{botta2016integration}.


\balance
\bibliographystyle{IEEEtran}
\bibliography{IEEEabrv,AoI}

\begin{thebibliography}{10}
\providecommand{\url}[1]{#1}
\csname url@samestyle\endcsname
\providecommand{\newblock}{\relax}
\providecommand{\bibinfo}[2]{#2}
\providecommand{\BIBentrySTDinterwordspacing}{\spaceskip=0pt\relax}
\providecommand{\BIBentryALTinterwordstretchfactor}{4}
\providecommand{\BIBentryALTinterwordspacing}{\spaceskip=\fontdimen2\font plus
\BIBentryALTinterwordstretchfactor\fontdimen3\font minus
  \fontdimen4\font\relax}
\providecommand{\BIBforeignlanguage}[2]{{%
\expandafter\ifx\csname l@#1\endcsname\relax
\typeout{** WARNING: IEEEtran.bst: No hyphenation pattern has been}%
\typeout{** loaded for the language `#1'. Using the pattern for}%
\typeout{** the default language instead.}%
\else
\language=\csname l@#1\endcsname
\fi
#2}}
\providecommand{\BIBdecl}{\relax}
\BIBdecl

\bibitem{yates2021age}
R.~D. Yates, Y.~Sun, D.~R. Brown, S.~K. Kaul, E.~Modiano, and S.~Ulukus, ``Age
  of information: An introduction and survey,'' \emph{{IEEE} J. Sel. Areas
  Commun.}, vol.~39, no.~5, pp. 1183--1210, May 2021.

\bibitem{kaul2012real}
S.~Kaul, R.~Yates, and M.~Gruteser, ``Real-time status: How often should one
  update?'' in \emph{Proc.\ IEEE Infocom}, 2012.

\bibitem{kaul2011minimizing}
S.~Kaul, M.~Gruteser, V.~Rai, and J.~Kenney, ``Minimizing age of information in
  vehicular networks,'' in \emph{Proc.\ IEEE SAHCN}, 2011, pp. 350--358.

\bibitem{wang2020preempt}
Y.~Wang, S.~Wu, L.~Yang, J.~Jiao, and Q.~Zhang, ``To preempt or not: Timely
  status update in the presence of non-trivial propagation delay,'' in
  \emph{Proc. IEEE VTC Fall}, 2020.

\bibitem{munari2022role}
A.~Munari and L.~Badia, ``The role of feedback in {AoI} optimization under
  limited transmission opportunities,'' in \emph{Proc.\ IEEE Globecom}, 2022.

\bibitem{crosara2022cost}
L.~Crosara and L.~Badia, ``Cost and correlation in strategic wireless sensing
  driven by age of information,'' in \emph{Proc.\ Eur. Wirel.}, 2022.

\bibitem{champati2021minimum}
J.~P. Champati, R.~R. Avula, T.~J. Oechtering, and J.~Gross, ``Minimum
  achievable peak age of information under service preemptions and request
  delay,'' \emph{{IEEE} J. Sel. Areas Commun.}, vol.~39, no.~5, pp. 1365--1379,
  May 2021.

\bibitem{munari}
A.~Munari, ``Modern random access: an age of information perspective on
  irregular repetition slotted {ALOHA},'' \emph{{IEEE} Trans. Commun.},
  vol.~69, no.~6, pp. 3572--3585, Jun. 2021.

\bibitem{badia2022game}
L.~Badia, A.~Zanella, and M.~Zorzi, ``Game theoretic analysis of age of
  information for slotted {ALOHA} access with capture,'' in \emph{Proc. IEEE
  Infocom Wkshps}, 2022.

\bibitem{yavascan2021analysis}
O.~T. Yavascan and E.~Uysal, ``Analysis of slotted {ALOHA} with an age
  threshold,'' \emph{{IEEE} J. Sel. Areas Commun.}, vol.~39, no.~5, pp.
  1456--1470, May 2021.

\bibitem{yates2012real}
R.~D. Yates and S.~Kaul, ``Real-time status updating: Multiple sources,'' in
  \emph{Proc.\ IEEE ISIT}, 2012, pp. 2666--2670.

\bibitem{bergson}
A.~Bergson, ``A reformulation of certain aspects of welfare economics,''
  \emph{Quart.\ J. Econ.}, vol.~52, no.~2, pp. 310--334, Feb. 1938.

\bibitem{moltafet2020average}
M.~Moltafet, M.~Leinonen, and M.~Codreanu, ``Average {AoI} in multi-source
  systems with source-aware packet management,'' \emph{{IEEE} Trans. Commun.},
  vol.~69, no.~2, pp. 1121--1133, Feb. 2020.

\bibitem{costa2016age}
M.~Costa, M.~Codreanu, and A.~Ephremides, ``On the age of information in status
  update systems with packet management,'' \emph{{IEEE} Trans. Inf. Theory},
  vol.~62, no.~4, pp. 1897--1910, Apr. 2016.

\bibitem{crosara2021stochastic}
L.~Crosara and L.~Badia, ``A stochastic model for age-of-information efficiency
  in {ARQ} systems with energy harvesting,'' in \emph{Proc.\ Eur. Wirel.},
  2021.

\bibitem{yates2018age}
R.~D. Yates and S.~K. Kaul, ``The age of information: Real-time status updating
  by multiple sources,'' \emph{{IEEE} Trans. Inf. Theory}, vol.~65, no.~3, pp.
  1807--1827, Mar. 2018.

\bibitem{talak2020age}
R.~Talak and E.~H. Modiano, ``Age-delay tradeoffs in queueing systems,''
  \emph{{IEEE} Trans. Inf. Theory}, vol.~67, no.~3, pp. 1743--1758, Mar. 2020.

\bibitem{nguyen2018information}
G.~D. Nguyen, S.~Kompella, C.~Kam, J.~E. Wieselthier, and A.~Ephremides,
  ``Information freshness over an interference channel: A game theoretic
  view,'' in \emph{Proc.\ IEEE Infocom}, 2018, pp. 908--916.

\bibitem{vadori2015jamming}
V.~Vadori, M.~Scalabrin, A.~V. Guglielmi, and L.~Badia, ``Jamming in underwater
  sensor networks as a {Bayesian} zero-sum game with position uncertainty,'' in
  \emph{Proc.\ IEEE Globecom}, 2015.

\bibitem{banerjee2022age}
S.~Banerjee and S.~Ulukus, ``Age of information in the presence of an
  adversary,'' in \emph{Proc.\ IEEE Infocom Wkshps}, 2022.

\bibitem{wang2019joint}
X.~Wang, L.~Liu, L.~Zhu, and T.~Tang, ``Joint security and {QoS} provisioning
  in train-centric {CBTC} systems under sybil attacks,'' \emph{{IEEE} Access},
  vol.~7, pp. 91\,169--91\,182, Jul. 2019.

\bibitem{jing2022joint}
T.~Jing, H.~Yu, X.~Wang, and Q.~Gao, ``Joint timeliness and security
  provisioning for enhancement of dependability in {Internet of Vehicle}
  system,'' \emph{Int.\ J. Distrib.\ Sens.\ Netw.}, vol.~18, no.~6, Jun. 2022.

\bibitem{asheralieva2022optimizing}
A.~Asheralieva and D.~Niyato, ``Optimizing age of information and security of
  the next-generation {Internet} of everything systems,'' \emph{{IEEE} Internet
  Things J.}, vol.~9, no.~20, pp. 20\,331--20\,351, Oct. 2022.

\bibitem{chen2020secure}
\BIBentryALTinterwordspacing
H.~Chen, Q.~Wang, P.~Mohapatra, and N.~Pappas, ``Secure status updates under
  eavesdropping: Age of information-based physical layer security metrics,''
  \emph{arXiv}, 2020. [Online]. Available:
  \url{https://arxiv.org/abs/2002.07340}
\BIBentrySTDinterwordspacing

\bibitem{nash1950bargaining}
J.~F. Nash, Jr, ``The bargaining problem,'' \emph{Econometrica}, vol.~18,
  no.~2, pp. 155--162, Apr. 1950.

\bibitem{zou2016survey}
Y.~Zou, J.~Zhu, X.~Wang, and L.~Hanzo, ``A survey on wireless security:
  Technical challenges, recent advances, and future trends,'' \emph{Proc.
  {IEEE}}, vol. 104, no.~9, pp. 1727--1765, Sep. 2016.

\bibitem{zhu2017secure}
Y.~Zhu, L.~Wang, K.-K. Wong, and R.~W. Heath, ``Secure communications in
  millimeter wave ad hoc networks,'' \emph{{IEEE} Trans. Wireless Commun.},
  vol.~16, no.~5, pp. 3205--3217, May 2017.

\bibitem{thanh2019efficient}
P.~Thanh, T.~Hoan, H.~Vu-Van, and I.~Koo, ``Efficient attack strategy for
  legitimate energy-powered eavesdropping in tactical cognitive radio
  networks,'' \emph{Wirel. Netw.}, vol.~25, pp. 3605--–3622, Feb. 2019.

\bibitem{principlesofcomm}
N.~Benvenuto and M.~Zorzi, \emph{Principles of communications networks and
  systems}.\hskip 1em plus 0.5em minus 0.4em\relax Wiley, 2006.

\bibitem{perin2021adversarial}
G.~Perin, A.~Buratto, N.~M. Anselmi, S.~Wagle, and L.~Badia, ``Adversarial
  jamming and catching games over {AWGN} channels with mobile players,'' in
  \emph{Proc.\ IEEE WiMob}, 2021, pp. 319--324.

\bibitem{botta2016integration}
A.~Botta, W.~De~Donato, V.~Persico, and A.~Pescap{\'e}, ``Integration of cloud
  computing and {I}nternet of things: a survey,'' \emph{Fut.\ Gen.\ Comp.\
  Syst.}, vol.~56, pp. 684--700, Mar. 2016.

\end{thebibliography}

\end{document}